\RequirePackage{lineno}
\documentclass[superscriptaddress,showpacs,aps,twocolumn,prl,floatfix]{revtex4}
\usepackage{amssymb}
\usepackage{amsmath}
\usepackage{graphicx,color}
\usepackage{bm}

\usepackage{ulem}           
\usepackage[english]{babel} 

  \newcommand{\Eq}[1]{Eq.~(\ref{#1})}

  \newcommand{\Fig}[1]{Fig.~\ref{#1}}

  \def\DE{E_{32}}

  \def\kT{k_B T}
  \def\eV{\,\mathrm{eV}}  
  \def\K{\,\mathrm{K}}    
  \def\Heff{\hat{H}_\mathrm{eff}} 
  \def\Simp{S_i}
  \def\TK{T_\mathrm{K}}

  \def\DeltaE{\Delta_\mathrm{xy}} 

  \def\DeltaZ{\Delta_z} 

  \def\Dast{T^\ast} 

  \def\Tastf{T^\ast_f}

\begin{document}
\title{Non-Fermi liquid behavior in transport through Co doped Au chains}

\author{S. Di Napoli}
\affiliation{Departamento de F\'{\i}sica de la Materia Condensada, CAC-CNEA, Avenida
General Paz 1499, (1650) San Mart\'{\i}n, Pcia. de Buenos Aires, Argentina and Consejo Nacional de Investigaciones Cient\'{\i}ficas
 y T\'ecnicas, CONICET, Buenos Aires, Argentina}

\author{A. Weichselbaum}
\affiliation{Ludwig Maximilians University and Arnold Sommerfeld Center for Theoretical Physics,
80333 Munich, Germany}

\author{P. Roura-Bas}
\affiliation{Departamento de F\'{\i}sica de la Materia Condensada, CAC-CNEA, Avenida
General Paz 1499, (1650) San Mart\'{\i}n, Pcia. de Buenos Aires, Argentina and Consejo Nacional de Investigaciones Cient\'{\i}ficas
 y T\'ecnicas, CONICET, Buenos Aires, Argentina}

\author{A. A. Aligia}
\affiliation{Centro At\'omico Bariloche and Instituto Balseiro, Comisi\'on Nacional de
Energ\'{\i}a At\'omica, 8400 Bariloche, Argentina}

\author{Y. Mokrousov}
\affiliation{ Peter Gr\"unberg Institut and Institute for Advanced Simulation, Forschungszentrum J\"ulich
 and JARA, 52425 J\"ulich, Germany}

\author{S. Bl\"{u}gel}
\affiliation{ Peter Gr\"unberg Institut and Institute for Advanced Simulation, Forschungszentrum J\"ulich
 and JARA, 52425 J\"ulich, Germany}

\begin{abstract}
We calculate the conductance as a function of temperature $G(T)$ through Au monoatomic chains 
containing one Co atom as a magnetic impurity, and 
connected to two conducting leads with a 4-fold symmetry axis.
Using the information derived from {\it ab initio} calculations, we
construct an effective model $\Heff$ that hybridizes a 3d$^7$ quadruplet at
the Co site with two 3d$^8$ triplets through the hopping of 5d$_{xz}$ and 5d$_{yz}$
electrons of Au. The quadruplet is split by
spin anisotropy due to spin-orbit coupling. Solving $\Heff$ with
the numerical renormalization group (NRG) 
we find that at low temperatures $G(T)=a-b \sqrt{T}$ and the ground state 
impurity entropy is $\ln(2)/2$, a behavior similar to the two-channel Kondo model.
Stretching the chain leads to a non Kondo phase, 
with the physics of the underscreened Kondo model at the quantum critical point.
\end{abstract}

\date{\today} 



\maketitle

\linenumbers

Transport properties in nanoscopic systems are being extensively studied due
to potential applications and their role in basic research. In
particular, the Kondo effect, in which the spin 1/2 of a molecule or a
quantum dot (QD) is screened by one channel of conduction electrons below a
characteristic temperature $T_{K}$, has been experimentally observed and the
conductance as a function of temperature $G(T)$ is in excellent agreement
with theory \cite{park,grobis}. In spite of the large effect of
correlations, these systems are Fermi liquids where, for example, the
conductance has the expected behavior $G(T)\simeq G(0)-aT^{2}$ for $T\ll T_{K}$ with $\TK$ the Kondo temperature
 and $a$ some constant.
More recently, the {\it underscreened} Kondo (UK) effect, 
leading to {\it singular} Fermi liquid physics \cite{meh,logan} 
has been observed, and
quantum phase transitions (QPTs) involving partially Kondo screened spin-1
molecular states were induced by 
externally controlled parameters 
\cite{roch,parks,serge}.

The {\it overscreened} Kondo effect, the simplest manifestation of which is
the two-channel Kondo (2CK) model, is even more interesting because
inelastic scattering persists even at vanishing temperatures and excitation
energies. Therefore the system is a non-Fermi liquid, exhibiting
fascinating low-energy properties \cite{bethe,zar,mit}. In particular, the
impurity contribution to the entropy is $\ln(2)/2$ and the conductance per
channel at low $T$ has the form $G(T)\simeq G_{0}/2\pm a\sqrt{T}$,
where $G_{0}$ is the conductance at zero temperature in the one-channel
case. However, experimental observations of 2CK physics have been elusive,
particularly due to its instability
against the asymmetry of
the channels \cite{mit}. For example, while 2CK physics is expected to take
place at the QPT in the model of two spin-1/2 Kondo impurities \cite{zar,mit}, 
inter-channel charge transfer spoils the quantum critical point (QCP) 
\cite{mit} and its observation in double QD systems \cite{bork}.

Oreg and Goldhaber-Gordon \cite{oreg} 
proposed to modify the simplest setup of the
one-electron transistor (a QD between two conducting leads) 
by adding a second large QD with level spacing
$\delta < \TK$, which acts as a second channel of conduction electrons,
with gate voltage fine tuned towards the QCP.
Following this proposal, the differential conductance as a function of bias
voltage $V_{b}$ was measured and a $\sqrt{V_{b}}$ behavior, characteristic of
2CK physics was observed for $eV_{b}>\kT$ \cite{potok}.

\begin{figure}[tbp]
\includegraphics[width=7.7cm]{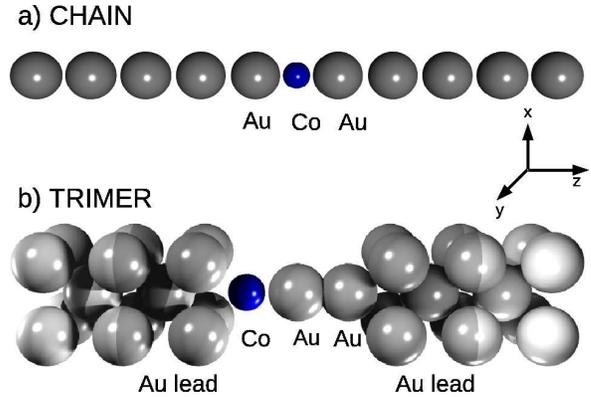}
\caption{(Color online) Structure of the systems studied. a) Au chain doped with Co.
b) trimer with one Co atom connected to BCC leads. The shaded region indicates the unit cell
used.}
\label{scheme}
\end{figure}

In this Letter, we show that 2CK physics is expected in
transport through Co doped Au chains connected to conducting leads with a
four-fold axis, as shown in \Fig{scheme}b. 
In this case, the mobile electrons of the two
channels are the 5d $xz$ and $yz$ electrons of Au.  
An advantage over the
proposal of Ref. \onlinecite{oreg}
is that both channels are related by
SU(2) symmetry in the effective model $\Heff$ and therefore
fine tuning is not necessary. Furthermore, by stretching the chain, it appears 
possible to pass through a QCP with {\it underscreened} Kondo physics to 
another phase with reduced conductance, without Kondo behavior.
To the best of our knowledge, our model has no analog to previous proposals for observation 
of 2CK physics \cite{cz}. 

The strategy of this Letter is twofold. First we perform
\textit{ab initio} calculations \cite{note1} to obtain an effective
low-energy Hamiltonian. This is followed by a detailed
analysis of the latter using analytical arguments together
with a numerical analysis based on the quasi exact
Numerical Renormalization Group (NRG)
\cite{wilson,bulla,wei2}.

\begin{figure}[tbp]
\includegraphics[width=\linewidth]{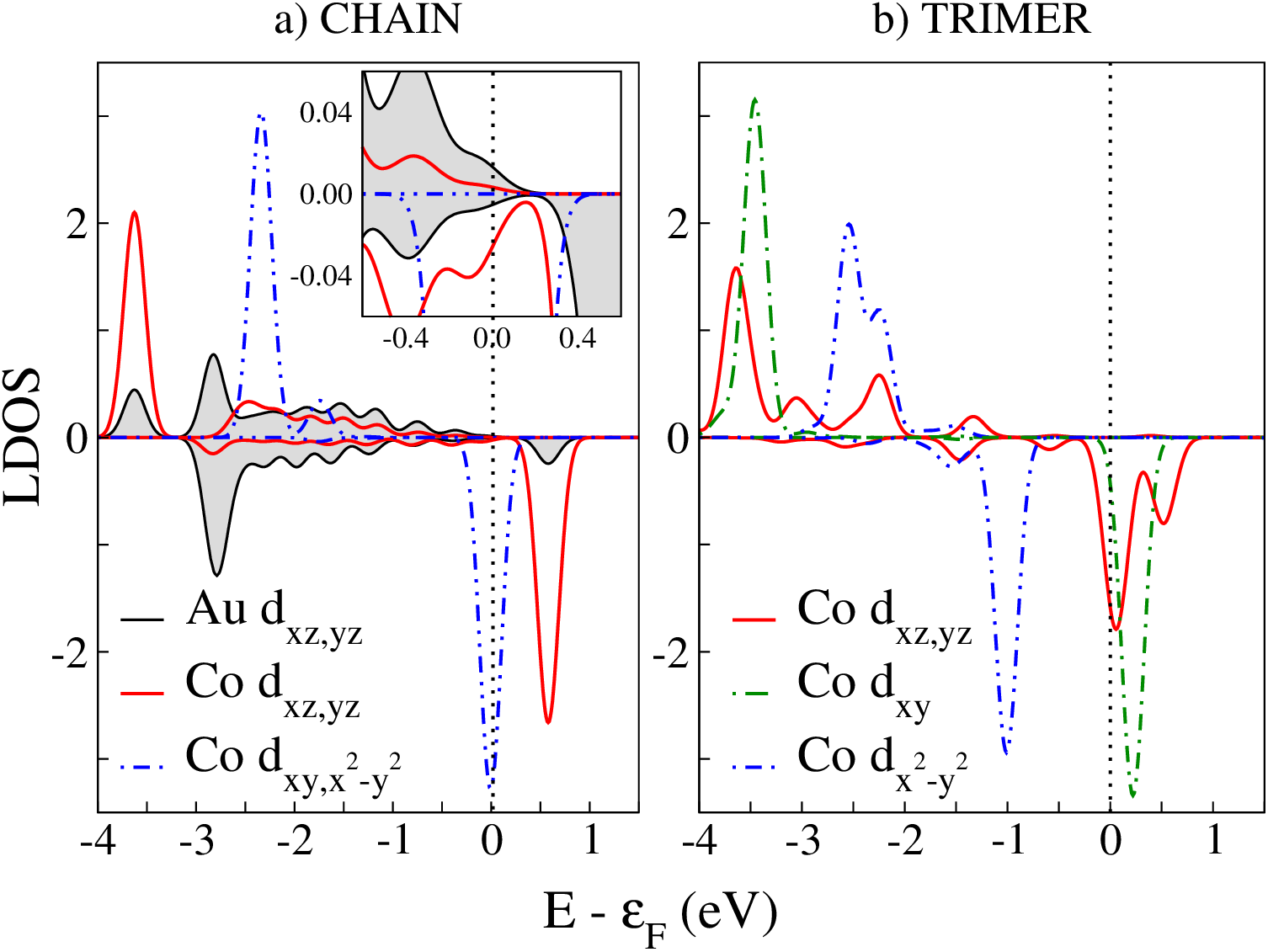}
\caption{(Color online) Spin-polarized spectral density of different orbitals in a) the linear chain 
Au$_{10}$Co depicted in \Fig{scheme}a, and b) the system of \Fig{scheme}b.}
\label{lda}
\end{figure}

Spin-polarized spectral densities of different orbitals 
are shown in \Fig{lda}. 
For the linear chain (\Fig{lda}a), we find that Co is in a 3d$^{7}$ configuration with total
spin 3/2. Two 3d holes reside in the $xz$ and $yz$ orbitals. The remaining
one is shared between the $x^{2}-y^{2}$ and $xy$ orbitals. The band
dispersion (not shown) indicates that these orbitals are strongly localized,
resulting in no Au orbitals with $x^{2}-y^{2}$ and $xy$ symmetry at the
Fermi energy $\epsilon_F$.  
We obtain that the top of the 5d $xz$ and $yz$ bands of Au are
slightly above $\epsilon_F$
(see the inset of \Fig{lda}a).
The same result
was obtained by previous \textit{ab initio} calculations of pure 
Au chains \cite{sclauzero12,zarechnaya08,yura05,delin03}. 
While transport experiments in Au chains indicate that these bands 
do not cross the Fermi level \cite{rego}, it has been found that
doping with oxygen
pushes these bands up establishing conduction through the 
5d$_{xz}$ and 5d$_{yz}$ electrons of Au \cite{sol12, thijssen06}. Application of a gate voltage 
can have a similar effect.

The degeneracy between $x^{2}-y^{2}$ and $xy$  Co orbitals is broken 
in a system as 
represented in \Fig{scheme}b. Previous calculations
for a Pt system with this geometry indicate that the $x^{2}-y^{2}$ and $xy$
orbitals are split by an energy $\DeltaE \approx 1.4\eV$, while the
remaining orbitals are unaffected \cite{thie} 
(for bulk gold, one has $\DeltaE \approx 1.0\eV$).
As shown in \Fig{lda}b, we obtain a similar splitting here. 
We have taken the Au-Au and Au-Co distances as 4.93 and 4.53 a.u,
respectively.
Choosing $\epsilon_F=0$, we obtain for the average
energies of the different
Co 3d orbitals 
$E_{x^{2}-y^{2}}=-1.0\eV$, $E_{3z^{2}-r^{2}}=-0.1\eV$ (not shown),
$E_{xy}=0.2\eV$ and $E_{xz}=E_{yz}=0.3\eV$. 
The occupations indicate some admixture of the 3d$^{7}$ quadruplets 
with 3d$^{8}$ triplets as described below.

An important ingredient of $\Heff$ is 
the splitting $D$ between 
Co spin states with projection $S_{z}=\pm 3/2$ from
those with $S_{z}=\pm 1/2$ originated from
spin-orbit coupling (SOC), calculated as described below.

The calculations above lead
us to the following 
effective Anderson Hamiltonian to describe the
low-energy physics,
\begin{eqnarray}
 && \Heff =
     \sum_{M_{3}} (E_{3}+\tfrac{D}{2}M_{3}^{2})
     |M_{3}\rangle \langle M_{3}|
   + E_{2}\sum_{\alpha M_{2}}|\alpha M_{2}\rangle \langle \alpha M_{2}|
     \notag \\
 &&+ \sum_{\alpha M_2 M_3} \sum_{\nu k \sigma} V_{\nu }
     \langle 1\tfrac{1}{2} M_{2}\sigma |\tfrac{3}{2} M_{3}\rangle
     \bigl( \hat{c}_{\nu k\alpha \sigma }^{\dagger }|\alpha M_{2}\rangle \langle M_{3}|
      + \mathrm{H.c.} \bigr)
     \notag \\
 &&+ \sum_{\nu k\alpha \sigma } \epsilon_{\nu k}
     \hat{c}^\dagger_{\nu k\alpha\sigma} \hat{c}^{\phantom{\dagger}}_{\nu k\alpha\sigma}
\text{.}\label{ham}
\end{eqnarray}
Here $E_{n}$, $M_{n}$ are the energies and spin projections
along the chain 
($z$-direction) of the states with $n\in \{2,3 \}$ holes in the 3d shell of Co. 
The charge-transfer energy is denoted by $\DE \equiv E_2 - E_3$.
For $n=3$, the state of maximum spin projection is $|\tfrac{3}{2}\rangle \equiv
\hat{d}_{xz\downarrow }^{\dagger } \hat{d}_{yz\downarrow }^{\dagger } \hat{d}_{xy\downarrow
}^{\dagger }|0\rangle $, where $|0\rangle $ is the full 3d$^{10}$ shell and 
$\hat{d}_{\beta \sigma }^{\dagger }$ creates a hole with symmetry $\beta $ and spin 
$\sigma $. Similarly, for two holes  $|\alpha 1\rangle
\equiv \hat{d}_{\alpha \downarrow
}|\tfrac{3}{2}\rangle $ ($\alpha \in\{ xz,yz\}$). The remaining relevant states at the Co
site can be constructed using the spin lowering operator.
The hopping amplitudes $V_{\nu }$  between the Co
atom and the states $\hat{c}_{\nu k\alpha \sigma }^{\dagger }$
are assumed  independent
of $k$, where
$\hat{c}_{\nu k\alpha \sigma }^{\dagger }$ creates a hole in the 5d band 
of Au states  with symmetry $\alpha $ at the
left ($\nu =L$) or right ($\nu =R$) of the Co site.
The hybridization of the Co atom to the mobile electrons
at either side \cite{meir} is given by $\Gamma _{\nu } \equiv 2\pi
\sum_{k}|V_{\nu }|^{2}\delta (\omega -\epsilon _{\nu k})$,
neglecting  the small dependence on $\omega$.

The Clebsch-Gordan coefficients  $\langle J_{2}\frac{1}{2}M_{2}\sigma
|J_{3}M_{3}\rangle $ lead to a spin SU(2) symmetric Hamiltonian for $D=0$ \cite{integ}.
For any $D$, the model exhibits channel SU(2) symmetry \cite{note2} which was also exploited 
in the NRG calculations below \cite{wei}. 

Depending on the value of $D$, the model exhibits
rich physics. For $D=0$,
the model is an SU(2) {\it underscreened} impurity Anderson model (two channels
with spin 1/2 cannot completely screen a spin 3/2). For one channel, this model has been solved 
exactly \cite{integ}. In the Kondo limit ($\DE \gg \Gamma$)  
it displays singular Fermi
liquid behavior \cite{meh,logan} observed in the conductance through molecules
containing Co$^+$ ions \cite{parks}. For two channels, similar physics is expected.
For sufficiently negative $D$,
 $|D|\gg \TK$, 
spin-flip
processes are inhibited, the Co spin projection $S_z$ fluctuates between 3/2 and 1
or between -3/2 and -1. The properties of the system are described by two independent
resonant levels \cite{bal}.
For large positive $D\gg \TK$, we find that a 
Schrieffer-Wolff 
transformation leads to an anisotropic
2CK model for the interaction between an effective spin 1/2 for the Co states with $S_z= \pm 1/2$
and the spin $s^{\alpha}$ of the mobile states that hybridize with the Co at the Fermi energy 
 
\begin{eqnarray}
\hat{H}_{\mathrm{2CK}} &=&
  J \sum_{\alpha} (\hat{S}_z \hat{s}_z^{\alpha}
+ 2 \hat{S}_x \hat{s}_x^{\alpha} + 2 \hat{S}_y \hat{s}_y^{\alpha})
\text{,}\label{h2ck}
\end{eqnarray}
where $J \simeq \tfrac{2}{3} (V_L^2 + V_R^2)/[\DE - (D/8)]$.
The critical behavior of this model is the same as
for the isotropic 2CK model \cite{aff}.
As a consequence, the system displays non-Fermi liquid behavior
due to {\it overscreening} of the impurity spin.

To calculate $D$ we exactly solved the $120 \times 120$
matrix of the Hamiltonian of the 3d$^7$ configuration,
as described in Ref. \cite{kroll},
including the effect of SOC via 
$\hat{H}_\mathrm{SOC} =
\lambda \sum_{i}\hat{\mathbf{l}}_{i} \cdot \hat{\mathbf{s}}_{i}$.
To gain insight on the effect of the splitting $\DeltaE$ between 3d $x^{2}-y^{2}$ and $xy$ orbitals,
we take the above mentioned \textit{ab initio} results for the energies of the system of \Fig{scheme}b, 
but assume $E_{xy}=-E_{x^{2}-y^{2}}=\DeltaE/2$ \cite{note3}.
For $\DeltaE=0$, we obtain
$D=-0.0426 \eV$,
indicating that the Co spin is oriented along the chain.
As discussed above, no particularly interesting physics is expected in this case.
However, as $\DeltaE$ increases, $D$ also increases and changes sign for
$\DeltaE \approx 0.8\eV$.
For $\DeltaE=1.2 \eV$ (corresponding to the setup of \Fig{scheme}b) we obtain $D=0.0017 \eV$.
One expects that
$\DeltaE$ is reduced by stretching the device and a QPT would take place if
$D=0$ is reached \cite{note4}.



The total resonant level width $\Gamma =\Gamma _{L}+\Gamma
_{R}=0.60\eV$, is
determined from the width of the peak of the $xz$ and $yz$ states above the
Fermi energy in the \textit{ab initio} calculations. From the position of
this peak, we take $\DE=0.3\eV$, choosing the origin of one-particle
energies at the Fermi energy $\epsilon _{F}=0$. For the numerical
calculations we take a band extending from $-W$ to $W$ with $W=5\eV$.
Repeating the \textit{ab initio} calculations for the system 
of \Fig{scheme}b either without the Au atoms in the chain or without
the left lead, we obtain $\Gamma _{R}=0.25\eV$, $\Gamma _{L}=0.34\eV$, 
leading to a factor $A=4 \Gamma _{L} \Gamma _{R}/\Gamma^2=0.977$ \cite{note}
for the conductance [see \Eq{cond}].

\begin{figure}[tbp]
\includegraphics[width=\linewidth]{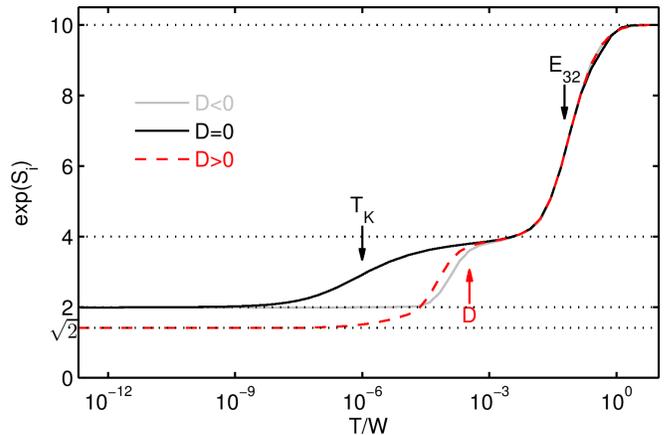}
\caption{(Color online) Impurity contribution to the entropy as a function of temperature for $D=0$ (full line black),
$D=-0.0017\eV$ (full line gray)
and $D=0.0017\eV$ (dashed line),
calculated using NRG [$\Lambda=3$,
$E_K=6.8$, 
keeping up to $6\,200$ multiplets ($15\,800$ states)].
}
\label{s0}
\end{figure}

\begin{figure}[tbp]
\includegraphics[width=\linewidth]{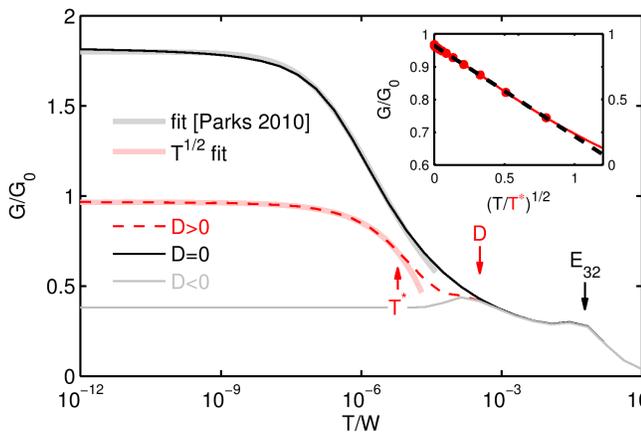}
\caption{(Color online) Conductance as a function of temperature
for $D=0$ (full black line), $D=-0.0017\eV$ (full gray line) 
and $D=0.0017\eV$ (dashed line), obtained
with NRG $[\Lambda=2$,
$E_K=6.5$,  
keeping up to 12\,300 multiplets
(31\,800 states)].
The inset shows a fit of the 
$D>0$ NRG data with $G=a-b \sqrt{T/\Dast}$ with
$\Dast/W=6\cdot 10^{-6}$.  
The resulting fit
is also replicated in the main panel (lower thick light line).
}
\label{g}
\end{figure}

The effective model in \Eq{ham} was simulated using NRG,
with the results presented in Figs.~\ref{s0} and \ref{g}
[discretization parameter $\Lambda$ and truncation energy
$E_K$ (in units of rescaled energies)
are specified in the captions]. 
In \Fig{s0} we show the impurity contribution to the entropy,
$\Simp$, 
averaging even and odd iterations.
At high temperatures, the quadruplet
and both triplets are equally populated and $\Simp \approx \ln(10)$.
As the temperature
is lowered below the charge-transfer energy $\DE$,
the system is in the local moment 
regime characterized by a local spin 3/2, and therefore a plateau appears with 
$\Simp \approx \ln(4)$. For $D=0$ (black solid line), as
the temperature is lowered below the Kondo temperature
$\TK/W \approx 5.2 \cdot 10^{-6}$, a partial screening
of the local spin takes place and
the ground state is a doublet with $\Simp = \ln(2)$. 
The same zero-temperature value is reached for negative $D$.

In contrast, for positive $D > \TK$, $S_i$
starts to fall more 
rapidly for $T < D$, and at low temperatures it saturates at the peculiar value 
$\ln(2)/2$ characteristic for 
2CK physics.
We have confirmed (not shown)
that the same low-temperature value is obtained for $D < \TK$.
From an analysis of the low-energy spectrum, we obtain that for 
$D \ll \TK$ the crossover occurs at an energy scale $\Dast \ll D$,
which roughly goes as $\Dast \approx \TK \exp[-(6\TK/D)^{1/2}]$. 
A similar crossover scale 
has been observed in a one-channel spin-1 UK model 
with anisotropy \cite{cor}.
The crossover scale $\Dast$ is present for any finite $D>0$.

Non-Fermi liquid behavior is also encountered
in the spectral density of localized electrons (not shown), which at zero temperature 
and at the Fermi energy in the Kondo limit is half the value expected
from the Friedel-Langreth sum rule \cite{mit}. These
peculiarities of the spectral density manifest 
themselves in measurable properties, such as the
total conductance through the system \cite{meir} 

\begin{equation}
G(T)=\frac{Ae}{h}\int d\omega\, \pi \Gamma \rho (\omega )\bigl( -\frac{\partial f%
}{\partial \omega }\bigr) ,  \label{cond}
\end{equation}
with 
$f(\omega )=1/(e^{\omega
/\kT}+1)$ the Fermi function, and $\rho (\omega )=$ $\sum_{\alpha \sigma
}\rho _{\alpha \sigma }(\omega )$ the impurity 
spectral density of states.

The results for $G(T)$ are presented in \Fig{g}.
For $D=0$ 
the conductance at zero temperature is
$G(0) \approx 1.8\,G_0$, where $G_0=2Ae^2/h$. This is nearly the ideal value $2G_0$ expected 
for perfect transmission with two channels. However, 
with increasing temperature 
the conductance falls more rapidly than for a usual Kondo model, 
reflecting the physics of the UK model 
\cite{meh,logan,roch,parks,serge}. 
We have obtained an excellent fit of the NRG data below $\TK$, using a phenomenological 
expression proposed for one-channel UK models 
\cite{parks,note5} (see fit Parks 2010 in \Fig{g}).

As the anisotropy $D$ is turned on,
the conductance $G(T)$
deviates from the behavior for $D=0$ for temperatures $T<|D|$.
In particular, for positive $D$
the conductance at zero temperature, $G(0)$, is nearly half the corresponding value for $D=0$.  
However, the most striking feature of $G(T)$ for $D>0$ is that it decreases with
increasing temperature following a $\sqrt{T}$ behavior, instead of the 
characteristic $T^2$ dependence of a Fermi liquid 
(as for $D<$).
Fitting the results for 
$T < 0.8\,T^\ast = 0.17 \K$, we obtain 
$G/G_0 \approx 0.96 - 0.47 \sqrt{T[K]}$
(see inset of \Fig{g}). 

The results above
demonstrate measurable non-Fermi-liquid properties
in the model. 
We have checked that the results are robust and do not change if for example anisotropy
in the triplets or more configurations of 3d Co states are added.
In real systems, an independent conduction channel
involving 6s states of Au and 4s and 3d$_{3z^2-r^2}$ states of Co contributes to the 
conductance, but this contribution is comparatively small because
it does not lead to a resonance near the Fermi energy (in contrast to the Kondo effect)
and more importantly, it does not alter the non-Fermi-liquid behavior.
The latter is only affected by a magnetic field or physical ingredients that break 
the channel SU(2) symmetry, as a splitting $\DeltaZ$ between Co 3d$_{xz}$
and 3d$_{yz}$ orbitals. This leads to a Fermi-liquid behavior below a temperature 
$\Tastf \approx (\nu \DeltaZ)^2 \TK$, where $\nu =1/2W \approx 0.1/$eV 
is the density of states at the Fermi 
energy \cite{sela}. For our parameters, $\TK \approx 2.6\cdot10^{-5}\eV$.
Therefore, for moderate $\DeltaZ$, $\Tastf$ is very small and 
non-Fermi-liquid physics is observable in a wide range of temperatures.
    
In summary, we have shown that it is possible to observe
non-Fermi-liquid behavior in transport through a Co doped Au
monoatomic chain, in a setup that is not far from currently accessible
technologies. It has the advantage over previous proposals in
similar nanoscopic systems that both channels are related by symmetry
and therefore, they are truly equivalent. Moderate symmetry-breaking
perturbations still allow to observe non-Fermi liquid physics in a wide range
of temperatures. Furthermore, by stretching the system it appears feasible
to induce a quantum phase transition to a phase with singular 
Fermi liquid behavior and increased conductance ($D=0$) and to another phase with 
reduced conductance ($D<0$).

This work was partially supported by PIP 11220080101821, PIP 00258 of
CONICET, and PICT R1776 of the ANPCyT, Argentina, and also by DFG (SFB-631, WE\-4819/1-1; A.W.), Germany.
Y.M. acknowledges funding under HGF YIG Programme VH-513.
P.R. is sponsored by Escuela de Ciencia y Tecnolog\'{\i}a, Universidad
Nacional de San Mart\'{\i}n. The authors are grateful to T. Costi and A. Thiess for important discussions. S. Di Napoli is grateful 
to A. M. Llois and M. A. Barral for useful discussions.


\begin{thebibliography}{99}
\bibitem{note2} While there are four channels of mobile electrons,
depending on symmetry ($xz$ or $yz$) or position with respect to the Co atom ($L$ or $R$), 
for each energy $\epsilon _{L k}= \epsilon _{R k'}$ for which there are states at the left and the right,
only the linear combination 
$V_{L}^{\phantom{\dagger}}
\hat{c}_{L k\alpha \sigma }^{\dagger }
+ V_{R}^{\phantom{\dagger}}
\hat{c}_{R k'\alpha \sigma }^{\dagger }$ 
hybridizes with the Co atom. Thus, the model 
effectively reduces to a two-channel impurity model,
with the Co atom playing the role of the impurity.
\bibitem{note3}
The calculation also requires the Coulomb integrals $F_i$ \cite{kroll}
and the SOC strength $\lambda$. From a fit of the
low energy spectra of late transition metal atoms, we take  
$F_{2}=0.16\eV$, $F_{4}=0.011\eV$, and $\lambda =0.08\eV$.
\bibitem{note4}
We find from our numerics, that the
anisotropy in \Eq{ham}, 
which solely acts on the states $|M_{3}\rangle$, 
can be generalized to a uniform anisotropy  
$\tfrac{D}{2} (S_{z,\mathrm{imp}}^\mathrm{tot})^2$ 
without much affecting the overall physics. 
Hence, for simplicity, a uniform anisotropy $D$ has been used.
\bibitem{note}
The total hybridization
$\Gamma \equiv \Gamma_L + \Gamma_R =0.59 \eV$, 
obtained by fitting the widths of the peaks
when either the part of the system to the right or to the left of the Co atom is cut off,
does not exactly coincide with the total $\Gamma=0.60 \eV$ 
obtained for the whole system. However, it is very close. We used $\Gamma$ in the NRG calculations 
and $\Gamma_L$ and $\Gamma_R$ to calculate the asymmetry factor  
$A=4 \Gamma _{L} \Gamma _{R}/(\Gamma _{L} + \Gamma _{R})^2$.
\end{thebibliography}
\end{document}